\PassOptionsToPackage{usenames,dvipsnames}{xcolor}
\documentclass[a4paper,11pt]{article}
\pdfoutput=1 

\usepackage{amsmath,amssymb}
\usepackage{booktabs}
\usepackage{braket}
\usepackage{fontenc}[T1]
\usepackage{graphicx}
\usepackage[utf8]{inputenc}
\usepackage{jheppub}
\usepackage{multirow}
\usepackage{placeins}
\usepackage{slashed}
\usepackage{xcolor}
\usepackage{xspace}

\newcommand{\para}{\parallel}

\renewcommand{\Re}{\operatorname{Re}}
\renewcommand{\Im}{\operatorname{Im}}

\newcommand{\dd}{\ensuremath{\text{d}}}

\newcommand{\EOS}{\texttt{EOS}\xspace}
\newcommand{\refapp}[1]{appendix~\ref{app:#1}}
\newcommand{\refeq}[1]{eq.~(\ref{eq:#1})}
\newcommand{\reffig}[1]{figure~\ref{fig:#1}}

\newcommand{\refsec}[1]{section~\ref{sec:#1}}
\newcommand{\reftab}[1]{table~\ref{tab:#1}}

\makeatletter

\def\dvd{\@ifstar\@@dvd\@dvd}
\newcommand{\@dvd}[1]{\textcolor{purple}{[\textbf{DvD:} #1]}}
\newcommand{\@@dvd}[1]{\textcolor{purple}{#1}}
\def\ak{\@ifstar\@@ak\@ak}
\newcommand{\@ak}[1]{\textcolor{ForestGreen}{[\textbf{AK:} #1]}}
\newcommand{\@@ak}[1]{\textcolor{ForestGreen}{#1}}
\def\pb{\@ifstar\@@pb\@pb}
\newcommand{\@pb}[1]{\textcolor{OrangeRed}{[\textbf{PB:} #1]}}
\newcommand{\@@pb}[1]{\textcolor{OrangeRed}{#1}}
\def\jt{\@ifstar\@@jt\@jt}
\newcommand{\@jt}[1]{\textcolor{blue}{[\textbf{JT:} #1]}}
\newcommand{\@@jt}[1]{\textcolor{blue}{#1}}
\makeatother

\allowdisplaybreaks

\title{Angular Analysis of \boldmath $\Lambda_b\to \Lambda_c (\to \Lambda \pi)\ell\bar\nu$}

\author{P.~B\"oer$^a$,}
\author{A.~Kokulu$^a$,}
\author{J.-N.~Toelstede$^{a,b}$,}
\author{D.~van~Dyk$^a$}
\affiliation[a]{Physik Department, Technische Universit\"at M\"unchen, James-Franck-Stra\ss{}e 1, D-85748 Garching, Germany}
\affiliation[b]{Max-Planck-Institute for Physics, F\"ohringer Ring 6, D-80805 Munich, Germany}

\preprint{EOS-2019-01, TUM-HEP-1213/19, MPP-2019-152}

\emailAdd{philipp.boeer@tum.de}
\emailAdd{ahmetkokulu@gmail.com}
\emailAdd{jan.toelstede@tum.de}
\emailAdd{danny.van.dyk@gmail.com}

\abstract{%
    We revisit the decay $\Lambda_b^0\to \Lambda_c^+ \ell^-\bar\nu$ ($\ell = e,\mu,\tau$)
    with a subsequent two-body decay $\Lambda_c^+ \to \Lambda^0 \pi^+$ in the Standard Model
    and in generic New Physics models. The decay's joint four-differential angular
    distribution can be expressed in terms of ten angular observables, assuming negligible
    polarization of the initial $\Lambda_b$ state. We present compact analytical
    results for all angular observables, which enables us to discuss their possible New Physics reach.
    We find that the decay at hand probes more and complementary independent combinations of Wilson coefficients
    compared to its mesonic counter parts $\bar{B}\to D^{(*)}\ell^-\bar\nu$.
    Our result for the angular distribution is at variance with some of the results on scalar-vector interference terms
    in the literature. We provide numerical estimates
    for all angular observables based on lattice-QCD results for the $\Lambda_b \to \Lambda_c$
    form factors and account for a recent measurement of the parity-violating parameter in
    $\Lambda_c^+\to \Lambda^0\pi^+$ decays by BESIII. A numerical implementation of our results is made
    publicly available as part of the \EOS software.
}

\newcommand{\op}[1]{\mathcal{O}_{#1}}

\newcommand{\wilson}[2][{}]{\mathcal{C}_{#2}^{#1}}
\newcommand{\Heff}{\mathcal{H}^\text{eff}}

\def\ALpe#1{{A_{\perp_{#1}}}}  \def\ALpa#1{{A_{\|_{#1}}}}

\def\ATpe#1{{A^T_{\perp_{#1}}}}

\begin{document}
\maketitle
\flushbottom

\section{Introduction}
\label{sec:intro}

Anomalous measurements \cite{Lees:2012xj,Lees:2013uzd,Huschle:2015rga,Aaij:2015yra,Hirose:2016wfn,Hirose:2017dxl,%
Aaij:2017uff,Aaij:2017deq,Abdesselam:2019dgh} in observables that probe Lepton Flavour Universality (LFU)
in $\bar{B}\to D^{(*)}\tau^-\bar\nu$ decays motivate careful reappraisal of the theoretical inputs
\cite{Neubert:1992pn,Ligeti:1993hw,Faller:2008tr,Fajfer:2012vx,Na:2015kha,Lattice:2015rga,Neubert:1992wq,Wang:2017jow,Gubernari:2018wyi}
entering the present Standard Model (SM) predictions of these decays.
At the same time, they also motivate further phenomenological analyses to uncover new observables
that complement the LFU probes, either in terms of independent systematic uncertainties, or in terms of independent constraints
on the origin of LFU effects. This work aims to achieve the latter by studying the four-differential decay rate of the cascade process
$\Lambda_b^0\to \Lambda_c^+(\to \Lambda^0 \pi^+)\ell^-\bar\nu$, with any charged lepton species $\ell = e,\mu,\tau$. To achieve
this goal, we work within the most general effective theory of local $b\to c\ell\bar\nu$ operators with left-handed neutrinos and
up to mass dimension six, which includes scalar and tensor interactions besides the SM contributions.
Our work extends earlier studies of the three-body decay $\Lambda_b^0\to\Lambda_c^+\ell^-\bar\nu$ \cite{Datta:2017aue},
and of the four-body decay $\Lambda_b^0\to \Lambda_c^+(\to \Lambda^0 \pi^+) \ell^-\bar\nu$~\cite{Shivashankara:2015cta,Gutsche:2015mxa}.
Our analysis follows closely previous works on the flavour-changing neutral-current decay
$\Lambda_b^0 \to \Lambda^0(\to p \pi^-)\ell^+\ell^-$~\cite{Boer:2014kda,Das:2018sms,Das:2018iap}, which
has a qualitatively similar angular distribution.\\

The remainder of this document is structured as follows. We present and discuss our analytical results for the four-body differential
decay rate and the ten angular observables arising from the latter in \refsec{analytical-results}.
In \refsec{numerical-results} we provide numerical results for the angular observables based on lattice QCD results for the full set of
$\Lambda_b\to \Lambda_c$ form factors~\cite{Meinel:2016dqj,Datta:2017aue} at mass dimension six, and based on our own average
of the parity violating parameter $\alpha$ in the secondary decay $\Lambda_c^+\to \Lambda^0\pi^+$~\cite{Ablikim:2019zwe}.
We conclude in \refsec{summary}.

\section{Analytical Results}
\label{sec:analytical-results}

We work within an effective field theory for semileptonic flavour-changing $|\Delta B| = |\Delta C| = 1$ transitions.
Its effective Hamiltonian can be expressed as
\begin{equation}
  \Heff = \frac{4 G_F}{\sqrt{2}} \tilde{V}_{cb} \left[
      \sum_{i}  \wilson{i} \op{i}
    \right] \,,
\end{equation}
where a basis of operators up to mass dimension six and with only left-handed neutrinos can be chosen as
\begin{equation}
    \begin{aligned}
        \op{V,L(R)}
            & \equiv \left[\bar{c} \gamma^\mu P_{L(R)} b\right]\,\left[\bar{\ell} \gamma_\mu P_L \nu_\ell\right]\,, &
        \op{S,L(R)}
            & \equiv \left[\bar{c} P_{L(R)} b\right]\,\left[\bar{\ell} P_L \nu_\ell\right]\,,                       \\
        \op{T}
            & \equiv \left[\bar{c} \sigma^{\mu\nu} b\right]\,\left[\bar{\ell} \sigma_{\mu\nu} P_L \nu_\ell\right]\,.
    \end{aligned}
    \label{eq:operator-basis}
\end{equation}
This approach is the standard way to account model-independently for the effects of New Physics (NP).
The SM corresponds to the parameter point $\wilson{V,L}=1$ and $\wilson{j} = 0$ for all remaining operators.
In the SM the parameter $\tilde{V}_{cb}$ coincides with the CKM matrix element $V_{cb}$. Beyond the
SM, it merely provides a common normalization for the effective operators. Throughout this work the value of
$\tilde{V}_{cb}$ is not relevant, since it cancels in the angular distribution. 
We note in passing that any interpretation of NP constraints in terms of the Wilson coefficients $\wilson{i}$
within the SM Effective Field Theory requires a careful interpretation of $\tilde{V}_{cb}$ \cite{Descotes-Genon:2018foz}.

\subsection{Angular Distribution}
\label{sec:analytical-results:ang-dist}

Following ref.~\cite{Boer:2014kda}, we express the kinematics of the four-body decay distribution
in terms of $q^2$, the square of the dilepton mass; $\cos\theta_\ell$, the helicity angle of the charged
lepton in the dilepton center-of-mass frame; $\cos\theta_{\Lambda_c}$, the helicity angle of the $\Lambda^0$
baryon in the $\Lambda^0 \pi^+$ center-of-mass frame; and $\phi$, the azimuthal angles between the
to decay planes. For more details, we refer to \refapp{kinematics}.
The fourfold differential distribution takes the form
\begin{align}
    \label{eq:KdGamma}
    K(q^2, \cos\theta_\ell, \cos \theta_{\Lambda_c}, \phi)
        & \equiv \frac{8\pi}{3} \, \frac{1}{\dd \Gamma / \dd q^2}\, \frac{\dd^4 \Gamma}{\dd q^2\,\dd \cos\theta_\ell\,\dd \cos \theta_{\Lambda_c} \,\dd \phi} \,.
\end{align}
The physical ranges of the kinematical variables are $m_\ell^2 \leq q^2 \leq (m_{\Lambda_b}-m_{\Lambda_c})^2$,
$\cos\theta_{\ell,\Lambda_c} \in \left[-1,1\right]$  and $\phi \in \left[ 0, 2\pi \right]$.
The distribution can be decomposed in terms of a set of trigonometric functions
\begin{equation}
\begin{aligned}
    \label{eq:angular-distribution}
    K(q^2, \cos\theta_\ell, \cos\theta_{\Lambda_c}, \phi)
        & = \big( K_{1ss} \sin^2\theta_\ell +\, K_{1cc} \cos^2\theta_\ell + K_{1c} \cos\theta_\ell\big)                    \\
        & + \big( K_{2ss} \sin^2\theta_\ell +\, K_{2cc} \cos^2\theta_\ell + K_{2c} \cos\theta_\ell\big) \cos \theta_{\Lambda_c} \\
        & + \big( K_{3sc}\sin\theta_\ell \cos\theta_\ell + K_{3s} \sin \theta_\ell\big) \sin \theta_{\Lambda_c} \sin\phi         \\
        & + \big( K_{4sc}\sin\theta_\ell \cos\theta_\ell + K_{4s} \sin\theta_\ell\big) \sin \theta_{\Lambda_c} \cos\phi         \,,
\end{aligned}
\end{equation}
giving rise to ten angular observables $K_i \equiv K_i(q^2)$.
Our choice of the normalization in \refeq{KdGamma} imposes an exact relation between two of the angular observables of the first row, $2K_{1ss} + K_{1cc} = 1$.
The structure of \refeq{angular-distribution} is imposed by angular momentum conservation \cite{Boer:2014kda}. It is generally compatible with the findings of
refs.~\cite{Shivashankara:2015cta,Gutsche:2015mxa}, that is, we find the same number of independent angular terms based on the
results for the angular distribution of refs.~\cite{Shivashankara:2015cta,Gutsche:2015mxa} as in \refeq{angular-distribution}.\\

In the presence of all five effective operators of \refeq{operator-basis}, we can express each angular observable as a sesquilinear
form of ten amplitudes
\begin{equation*}
    \lbrace A_{\lambda_m}, A^T_{\lambda_m} \rbrace = \lbrace A_{\perp_t}, A_{\perp_0}, A_{\perp_1}, A_{\para_t}, A_{\para_0}, A_{\para_1}, A^T_{\perp_0}, A^T_{\perp_1}, A^T_{\para_0}, A^T_{\para_1}\rbrace\,. 
\end{equation*}
Here $\lambda=\perp,\parallel$ denotes the transversity state; $m=t$ denotes time-like dilepton state; $m=0,1$ denotes the magnitude
of the $z$-component of the dilepton angular momentum in a vector dilepton state; and the $T$ superscript indicates that an amplitude
arises only in the presence of tensor operators. For the first row of \refeq{angular-distribution} we find:
\begin{align}
    \label{eq:K1ss}
    \frac{\dd \Gamma}{\dd q^2}\,
    K_{1ss}
         & = \frac{1}{4}\bigg[\left(1+\frac{m_\ell^2}{q^2}\right)|\ALpe1|^2
          + 2 |\ALpe0|^2  + 2 \frac{m_\ell^2}{q^2} |\ALpe{t}|^2 + (\perp \, \leftrightarrow \, \para)  \bigg] \\
    \nonumber
         & \quad - 2 \frac{m_\ell}{\sqrt{q^2}} \Re \bigg\lbrace A^T_{\perp_0} A^{*}_{\perp_0} + A^T_{\perp_1} A^{*}_{\perp_1} + (\perp \, \leftrightarrow \, \para) \bigg\rbrace \\
    \nonumber
        & \quad + \left[ \left( 1 + \frac{m_\ell^2}{q^2} \right) |\ATpe1|^2 + 2 \frac{m_\ell^2}{q^2} |\ATpe0|^2 +(\perp \, \leftrightarrow \, \para) \right] \,,\\
    \label{eq:K1cc}
    \frac{\dd \Gamma}{\dd q^2}\,
    K_{1cc}
        & = \frac{1}{2}\bigg[ |\ALpe1|^2 +  \frac{m_\ell^2}{q^2} \big( |\ALpe0|^2 + |\ALpe{t}|^2 \big) + (\perp \, \leftrightarrow \, \para) \bigg] \\
    \nonumber
        & \quad - 2 \frac{m_\ell}{\sqrt{q^2}} \Re \bigg\lbrace A^T_{\perp_0} A^{*}_{\perp_0} + A^T_{\perp_1} A^{*}_{\perp_1} + (\perp \, \leftrightarrow \, \para) \bigg\rbrace \\
    \nonumber
        & \quad + 2 \left[ |A^T_{\perp_0}|^2 +  \frac{m_\ell^2}{q^2} |A^T_{\perp_1}|^2 + (\perp \, \leftrightarrow \, \para)\right] \,, \\
    \label{eq:K1c}
    \frac{\dd \Gamma}{\dd q^2}\,
    K_{1c}
        & = \Re \left\lbrace A_{\perp_1}A^{*}_{\|_1} + \frac{m_\ell^2}{q^2} \big( A_{\perp_0} A^{*}_{\perp_t} + A_{\para_0} A^{*}_{\para_t} \big) \right\rbrace \\
    \nonumber
        & \quad - 2 \frac{m_\ell}{\sqrt{q^2}} \Re \left\lbrace A^T_{\perp_0} A^{*}_{\perp_t} + A^T_{\perp_1} A^{*}_{\para_1} +  (\perp \, \leftrightarrow \, \para) \right\rbrace \\
    \nonumber
        & \quad + 4 \frac{m_\ell^2}{q^2} \Re \left\lbrace A^T_{\perp_1} A^{T*}_{\para_1} \right\rbrace \,.
\end{align}
For the second row of \refeq{angular-distribution} we find:
\begin{align}
    \label{eq:K2ss}
    \frac{\dd \Gamma}{\dd q^2}\,
    K_{2ss}
        & = \frac{\alpha}{2} \Re \left\lbrace \bigg(1+\frac{m_\ell^2}{q^2}\bigg) A_{\perp_1}A^{*}_{\|_1} + 2 A_{\perp_0}A^{*}_{\|_0} +  2\frac{m_\ell^2}{q^2} A_{\perp_t} A^{*}_{\para_t} \right\rbrace \\
    \nonumber
        & \quad - 2 \, \alpha \frac{m_\ell}{\sqrt{q^2}} \Re \left\lbrace A^T_{\perp_0} A^{*}_{\para_0} + A^T_{\perp_1} A^{*}_{\para_1} + (\perp \, \leftrightarrow \, \para) \right\rbrace \\
    \nonumber
        & \quad + 2\,  \alpha \Re \left\lbrace \Big(1+\frac{m_\ell^2}{q^2} \Big) A^T_{\perp_1} A^{T*}_{\para_1} + 2 \frac{m_\ell^2}{q^2} A^T_{\perp_0} A^{T*}_{\para_0} \right\rbrace \,, \\
    \label{eq:K2cc}
    \frac{\dd \Gamma}{\dd q^2}\,
    K_{2cc}
        & = \alpha \Re \left\lbrace A_{\perp_1} A^{*}_{\para_1} + \frac{m_\ell^2}{q^2}\big( A_{\perp_0} A^{*}_{\para_0} + A_{\perp_t} A^{*}_{\para_t} \big)  \right\rbrace \\
    \nonumber
        & \quad - 2 \, \alpha \frac{m_\ell}{\sqrt{q^2}} \Re \left\lbrace A^T_{\perp_0} A^{*}_{\para_0} + A^T_{\perp_1} A^{*}_{\para_1} + (\perp \, \leftrightarrow \, \para) \right\rbrace \\
    \nonumber
        & \quad + 4 \, \alpha \Re \left\lbrace A^T_{\perp_0} A^{T*}_{\para_0} + \frac{m_\ell^2}{q^2} A^T_{\perp_1} A^{T*}_{\para_1} \right\rbrace \,, \\
    \label{eq:K2c}
    \frac{\dd \Gamma}{\dd q^2}\,
    K_{2c}
        & = \frac{\alpha}{2}  \left[ |\ALpe1|^2 + 2\frac{m_\ell^2}{q^2} \Re \left\lbrace A_{\perp_0} A^{*}_{\para_t}  \right\rbrace  + (\perp \, \leftrightarrow \,  \para) \right] \\
    \nonumber
        & \quad - 2\, \alpha \frac{m_\ell}{\sqrt{q^2}} \Re \left\lbrace A^T_{\perp_0} A^{*}_{\para_t} +  A^T_{\perp_1} A^{*}_{\perp_1} + (\perp\,  \leftrightarrow \, \para) \right\rbrace \\
    \nonumber
        & \quad + 2\, \alpha  \frac{m_\ell^2}{q^2} \bigg[ |A^T_{\perp_1}|^2 + (\perp \, \leftrightarrow\, \para) \bigg] \,. 
\end{align}
For the third row of \refeq{angular-distribution} we find:
\begin{align}
    \label{eq:K3sc}
    \frac{\dd \Gamma}{\dd q^2}\,
    K_{3sc}
         = &- \frac{\alpha}{\sqrt{2}} \bigg( 1-\frac{m_\ell^2}{q^2} \bigg) \Im \left\lbrace A_{\perp_0} A^{*}_{\perp_1} - (\perp\,  \leftrightarrow \, \para) \right\rbrace \\
    \nonumber
        & + 2\, \sqrt{2}\, \alpha \bigg(1-\frac{m_\ell^2}{q^2}\bigg) \Im \left\lbrace A^T_{\perp_0} A^{T*}_{\perp_1} - (\perp\,  \leftrightarrow \, \para) \right\rbrace \,, \\
    \label{eq:K3s}
    \frac{\dd \Gamma}{\dd q^2}\,
    K_{3s}
        = &-\frac{\alpha}{\sqrt{2}} \Im \left\lbrace A_{\perp_0} A^{*}_{\para_1} + \frac{m_\ell^2}{q^2} A_{\perp_1} A^{*}_{\perp_t} - (\perp\,  \leftrightarrow \, \para) \right\rbrace \\ 
    \nonumber
        &+ \sqrt{2} \, \alpha \frac{m_\ell}{\sqrt{q^2}} \Im \left\lbrace A^T_{\perp_0} A^{*}_{\para_1} + A^T_{\perp_1} A^{*}_{\para_0} + A^T_{\perp_1} A^{*}_{\perp_t} - (\perp\,  \leftrightarrow \, \para) \right\rbrace \\
    \nonumber 
        & - 2\, \sqrt{2} \, \alpha \frac{m_\ell^2}{q^2} \Im \left\lbrace A^T_{\perp_0} A^{T*}_{\para_1} - (\perp\,  \leftrightarrow \, \para) \right\rbrace \,.
\end{align}
For the fourth row of \refeq{angular-distribution} we find:
\begin{align}
    \label{eq:K4sc}
    \frac{\dd \Gamma}{\dd q^2}\,
    K_{4sc}
         = &- \frac{\alpha}{\sqrt{2}} \bigg(1-\frac{m_\ell^2}{q^2} \bigg) \Re \left\lbrace A_{\perp_0} A^{*}_{\para_1} - (\perp\,  \leftrightarrow \, \para) \right\rbrace \\
    \nonumber
        & + 2\, \sqrt{2} \, \alpha \bigg(1-\frac{m_\ell^2}{q^2} \bigg) \Re \left\lbrace A^T_{\perp_0} A^{T*}_{\para_1} - (\perp\,  \leftrightarrow \, \para) \right\rbrace \,, \\
    \label{eq:K4s}
    \frac{\dd \Gamma}{\dd q^2}\,
    K_{4s} 
         = &-\frac{\alpha}{\sqrt{2}} \Re \left\lbrace A_{\perp_0} A^{*}_{\perp_1} +\frac{m_\ell^2}{q^2} A_{\perp_1} A^{*}_{\para_t} - (\perp\,  \leftrightarrow \, \para) \right\rbrace  \\ 
    \nonumber
        & + \sqrt{2}\, \alpha \frac{m_\ell}{\sqrt{q^2}} \Re \left\lbrace A^T_{\perp_0} A^{*}_{\perp_1} + A^T_{\perp_1} A^{*}_{\perp_0} + A^T_{\perp_1} A^{*}_{\para_t} - (\perp\,  \leftrightarrow \, \para) \right\rbrace \\ 
    \nonumber 
        & - 2\, \sqrt{2}\, \alpha \frac{m_\ell^2}{q^2} \Re \left\lbrace A^T_{\perp_0} A^{T*}_{\perp_1} - (\perp\,  \leftrightarrow \, \para) \right\rbrace\,.
\end{align}
Comparing our results to the literature, our findings are summarized as follows:
\begin{itemize}
    \item We find complete agreement with the results of ref.~\cite{Datta:2017aue}, which discusses the non-cascade decay
        in the presence of all dimension-six operators.
    \item Our findings are at variance with the results of ref.~\cite{Shivashankara:2015cta}, which discusses
        the cascasde decay for all dimension-six operators \emph{except} for the tensor operator. We find
        interference terms of the type $V\times S$ and $A \times P$
        in our angular observables $K_{3s}$ and $K_{4s}$ that are absent from the corresponding term
        $C_4^{int}$ of ref.~\cite[arXiv v4]{Shivashankara:2015cta}.
        We believe our results to be correct, since they pass a crucial cross check: the interference
        between vector and scalar operators can be completely described by a shift of the time-like transversity
        amplitudes $\ALpe{t}$ and $\ALpa{t}$. This shift is consistent with a Ward-like identity for the
        $\Lambda_b\to \Lambda_c$ matrix elements:
        \begin{equation}
         \bra{\Lambda_c}  \bar{c} (\gamma_5)\, b \ket{\Lambda_b} = \frac{q^\mu}{m_b \mp m_c} \bra{\Lambda_c} \bar{c} (\gamma_5) \gamma_\mu b \ket{\Lambda_b} \,.
        \end{equation}
        The term $C_4^{int}$ in eq. (19) of  ref.~\cite{Shivashankara:2015cta} cannot
        be expressed through this shift. Additionally, we find a lack of an overall multiplicative factor
        $\mathcal{B}(\Lambda_c^+\to \Lambda^0 \pi^+)$ from the secondary decay. Furthermore, we find
        agreement only by redefining the angles $\theta_\ell \to \pi-\theta_\ell$, $\theta_{\Lambda_c} \to \theta_s$ and $\phi \to -\chi$.
    \item Comparing our results to the SM results for the fourfold distribution in ref. \cite{Gutsche:2015mxa}, we find complete agreement when
        redefining the angles as $\theta_\ell \to \pi-\theta_\ell$, $\theta_{\Lambda_c} \to \theta_s$ and $\phi \to -\chi$.
        We note the absence of any terms corresponding to the angular observables $K_{3sc}$ and $K_{3s}$, which vanish in the SM.
\end{itemize}

In the definition of seven out of the ten observables, we explicitly factor out the quantity $\alpha \equiv \alpha(\Lambda_c^+\to \Lambda^0\pi^+)$,
which is an angular asymmetry parameter arising in $\Lambda_c^+ \to \Lambda^0 \pi^+$ decays. It emerges as the hadronic matrix element of the parity-violating
weak decay of the $\Lambda_c^+$.
The present world average of this parameter by the Particle Data Group (PDG) \cite{Tanabashi:2018oca}  reads:
\begin{equation}
    \alpha^\text{PDG} = -0.91 \pm 0.15\,.
\end{equation}
It includes measurements by the CLEO \cite{Avery:1990ya} and ARGUS \cite{Albrecht:1991vs} experiments, in addition to
the statistically dominating measurements by the CLEO-2 \cite{Bishai:1995gp} $e^+e^-$ collider experiment,
\begin{equation}
    \alpha^\text{CLEO-2} = -0.94 \pm 0.21(\text{stat.}) \pm 0.12(\text{syst.})\,,
\end{equation}
and by the FOCUS (Fermilab E-831) \cite{Link:2005ft}  fixed target experiment,
\begin{equation}
    \alpha^\text{FOCUS} = -0.78 \pm 0.16(\text{stat.}) \pm 0.13(\text{syst.})\,.
\end{equation}
With the recent measurement of $\alpha$ by the BESIII collaboration in $e^+e^-$ collisions,
\begin{equation}
    \alpha^\text{BESIII} = -0.80 \pm 0.11(\text{stat.}) \pm 0.02(\text{syst.})\,,
\end{equation}
we compute our own average including also the CLEO-2 and FOCUS results. We obtain:
\begin{equation}
    \alpha = -0.82 \pm 0.09
\end{equation}
In the following we will use this average in lieu of an updated world average by the PDG.

\subsection{Transversity Amplitudes}

The transversity amplitudes arising in the angular observables $K_i$ are further decomposed
into $\Lambda_b \to \Lambda_c$ helicity form factors, Wilson coefficients, and kinematical factors.
Equations of motion for the $b$ and $c$-quark fields relate the hadronic matrix elements of scalar
(pseudo-scalar) currents to timelike vector (axial-vector) form factors. In the absence of the tensor operator
we find six independent transversity amplitudes:
\begin{align}
    A_{\perp_1} & = -2 \mathcal{N} \sqrt{s_-} \, f_\perp^V(q^2) \, \wilson{V,+}
                        \equiv F_{\perp_1} \, \wilson{V,+} \,, \\
    A_{\para_1} & = +2 \mathcal{N} \sqrt{s_+} \, f_\perp^A(q^2) \, \wilson{V,-}
                        \equiv F_{\para_1} \, \wilson{V,-} \,, \\
    A_{\perp_0} & = +\sqrt{2} \mathcal{N} \sqrt{s_-} \frac{m_{\Lambda_b}+m_{\Lambda_c}}{\sqrt{q^2}} \, f_0^V(q^2) \, \wilson{V,+}
                        \equiv F_{\perp_0} \, \wilson{V,+} \,, \\
    A_{\para_0} & = -\sqrt{2} \mathcal{N} \sqrt{s_+} \frac{m_{\Lambda_b}-m_{\Lambda_c}}{\sqrt{q^2}} \, f_0^A(q^2) \, \wilson{V,-}
                        \equiv F_{\para_0} \, \wilson{V,-} \,, \\
    A_{\perp_t} & = +\sqrt{2} \mathcal{N} \sqrt{s_+} \frac{m_{\Lambda_b}-m_{\Lambda_c}}{m_\ell} \, f_t^V(q^2) \,
                      \Big[ \frac{m_\ell}{\sqrt{q^2}} \, \wilson{V,+} + \frac{\sqrt{q^2}}{m_b - m_c} \, \wilson{S,+} \Big] \\
    \nonumber
                  & \equiv F_{\perp_t} \, \wilson{V,+} + \frac{\sqrt{q^2}}{m_\ell}F^S_{\perp} \, \wilson{S,+} \,, \\
    A_{\para_t} & = -\sqrt{2} \mathcal{N} \sqrt{s_-} \frac{m_{\Lambda_b}+m_{\Lambda_c}}{m_\ell} \, f_t^A(q^2) \,
                      \Big[ \frac{m_\ell}{\sqrt{q^2}} \, \wilson{V,-} - \frac{\sqrt{q^2}}{m_b + m_c} \, \wilson{S,-} \Big] \\
    \nonumber
                  & \equiv F_{\para_t} \, \wilson{V,-} + \frac{\sqrt{q^2}}{m_\ell}F^S_{\para} \, \wilson{S,-} \,,
\end{align}
with
\begin{equation}
    s_\pm \equiv \big(m_{\Lambda_b} \pm m_{\Lambda_c}\big)^2 - q^2\,.
\end{equation}
We note that the timelike leptonic polarization state requires a non-vanishing lepton mass. Hence, the transversity
amplitudes $A_{\perp_t}$ and $A_{\para_t}$ always enter observabes with a factor $m_\ell$ such that all physical observables
are well-defined in the limit $m_\ell \to 0$.
In the above we introduce a phenomenologically useful basis of effective form factors $F_{\lambda_m}$ and $F^S_\lambda$,
and abbreviate common linear combinations of the Wilson coefficients:
\begin{align}
    \wilson{V,\pm} & \equiv \wilson{V,L} \pm \wilson{V,R}\,, &
    \wilson{S,\pm} & \equiv \wilson{S,L} \pm \wilson{S,R}\,.
\end{align}
We also introduce an overall normalization factor
\begin{equation}
    \mathcal{N} \equiv G_F \, \tilde{V}_{cb} \left( 1 - \frac{m_\ell^2}{q^2} \right) \sqrt{\frac{ q^2 \sqrt{\lambda(m_{\Lambda_b}^2,m_{\Lambda_c}^2,q^2)}}{3 \times 2^{7} \pi^3 m_{\Lambda_b}^3} \times \mathcal{B}(\Lambda_c^+\to \Lambda^0 \pi^+)} \,,
\end{equation}
with the K\"all\'en function $\lambda(a, b, c) = a^2 + b^2 + c^2 - 2 ab - 2ac - 2 bc$.

In the presence of the tensor operator we find four additional transversity amplitudes and hence four additional
effective form factors $F^T_{\lambda_m}$:
\begin{align}
    A^T_{\para_0} & = -2\sqrt{2} \mathcal{N} \sqrt{s_+} \, \wilson{T} f_0^{T5}(q^2) \equiv F^T_{\para_0} \, \wilson{T}\,, \\
    A^T_{\perp_0} & = -2\sqrt{2} \mathcal{N} \sqrt{s_-} \, \wilson{T} f_0^{T}(q^2)  \equiv F^T_{\perp_0} \, \wilson{T}\,, \\
    A^T_{\para_1} & = +4 \mathcal{N} \frac{m_{\Lambda_b}-m_{\Lambda_c}}{\sqrt{q^2}} \sqrt{s_+} \, \wilson{T} f_\perp^{T5}(q^2) \equiv F^T_{\para_1} \, \wilson{T} \,, \\
    A^T_{\perp_1} & = +4 \mathcal{N} \frac{m_{\Lambda_b}+m_{\Lambda_c}}{\sqrt{q^2}} \sqrt{s_-} \, \wilson{T} f_\perp^{T}(q^2)  \equiv F^T_{\perp_1} \, \wilson{T} \,.
\end{align}

\subsection{Phenomenology}

To understand the NP reach of any of the angular observables or their combinations, it is instrumental to understand the 
sesquilinear combinations of Wilson coefficients entering the observables. To facilitate this understanding we define
five real-valued and ten complex-valued quantities $\sigma$:
\begin{equation}
\begin{aligned}
    \sigma_{V,1}^\pm  & =  \frac{1}{2} |\wilson{V,\pm}|^2 \,, &
    \sigma_{V,2}      & = -\wilson{V,-} \wilson[*]{V,+}   \,, \\
    \sigma_{S,1}^\pm  & =  \frac{1}{2} |\wilson{S,\pm}|^2 \,, &
    \sigma_{S,2}      & = -\wilson{S,-} \wilson[*]{S,+}   \,, \\
    \sigma_{VS,1}^\pm & =  \frac{1}{2}  \wilson{V,\pm} \wilson[*]{S,\pm} \, &
    \sigma_{VS,2}     & = -\wilson{V,-} \wilson[*]{S,+}   \,, \\
                      &                                       &
    \sigma_{SV,2}     & = -\wilson{S,-} \wilson[*]{V,+}   \,, \\
    \sigma_{T,1}      & =  \frac{1}{2} |\wilson{T}|^2     \,, \\
    \sigma_{VT,1}^\pm & =  \frac{1}{2} \wilson{V,\pm} \wilson[*]{T} \,, \\
    \sigma_{ST,1}^\pm & =  \frac{1}{2} \wilson{S,\pm} \wilson[*]{T} \,.
\end{aligned}
\end{equation}
This constitutes the complete set of sesquilinear combinations of the Wilson coefficients,
and corresponds to 25 real-valued degrees of freedom.
In terms of these combinations and effective form factors, we find in the limit $m_\ell \to 0$
\begin{align}
    \label{eq:K1ssm0}
    \frac{\dd\Gamma}{\dd q^2}
    K_{1ss}
        & = \frac{1}{2}\sigma_{V,1}^+ \bigg(2\,|F_{\perp_0}|^2+|F_{\perp_1}|^2 \bigg) +\frac{1}{2}\sigma_{V,1}^- \bigg(2\,|F_{\para_0}|^2+|F_{\para_1}|^2 \bigg) \\ 
        \nonumber
        &\quad + 4\bigg( \sigma_{S,1}^+|F^S_{\perp}|^2+\sigma_{S,1}^-|F^S_\para|^2 \bigg) + 2\,\sigma_{T,1} \bigg( |F^T_{\perp_1}|^2+|F^T_{\para_1}|^2 \bigg) \,, \\
    \label{eq:K1ccm0}
    \frac{\dd\Gamma}{\dd q^2}
    K_{1cc}
        & = \sigma_{V,1}^+|F_{\perp_1}|^2+\sigma_{V,1}^-|F_{\para_1}|^2 +\sigma_{S,1}^+|F^S_{\perp}|^2+ \sigma_{S,1}^-|F^S_{\para}|^2 \\
        \nonumber
        & \quad +4\,\sigma_{T,1} \bigg( |F^T_{\perp_0}|^2+|F^T_{\para_0}|^2 \bigg) \,, \\
    \label{eq:K1cm0}
    \frac{\dd\Gamma}{\dd q^2}
    K_{1c} 
        & = - \Re \left\lbrace \sigma_{V,2} F_{\para_1} F^*_{\perp_1} + 4\,\sigma_{ST,1}^+ F^S_\perp F^{T*}_{\perp_0} + 4\, \sigma_{ST,1}^- F^S_{\para} F^{T*}_{\para_0} \right\rbrace  \,, \\
    \label{eq:K2ssm0}
    \frac{\dd\Gamma}{\dd q^2}
    K_{2ss} 
        & = -\frac{\alpha}{2} \Re \left\lbrace \sigma_{V,2} \bigg( F_{\para_1} F^*_{\perp_1} + 2\,F_{\para_0} F^*_{\perp_0} \bigg) + 2\,\sigma_{S,2} F^S_{\para} F^{S*}_{\perp} - 8\,\sigma_{T,1} F^T_{\para_1} F^{T*}_{\perp_1} \right\rbrace \, \\
    \label{eq:K2ccm0}
    \frac{\dd\Gamma}{\dd q^2}
    K_{2cc}
        & = -\alpha \Re \left\lbrace \sigma_{V,2} F_{\para_1} F^*_{\perp_1} + \sigma_{S,2} F^S_{\para} F^{S*}_{\perp} -8\,\alpha\,\sigma_{T,1} F^T_{\para_0} F^{T*}_{\perp_0} \right\rbrace  \,,\\
    \label{eq:K2cm0}
    \frac{\dd\Gamma}{\dd q^2}
    K_{2c} 
        & = \alpha \bigg( \sigma_{V,1}^+ |F_{\perp_1}|^2 + \sigma_{V,1}^- |F_{\para_1}|^2 \bigg) -4\, \alpha \Re \left\lbrace \sigma_{ST,1}^+ F^S_{\perp} F^{T*}_{\para_0} + \sigma_{ST,1}^- F^S_{\para} F^{T*}_{\perp_0} \right\rbrace  \,,\\
    \label{eq:K3scm0}
    \frac{\dd\Gamma}{\dd q^2}
    K_{3sc}
        & = -\sqrt{2}\,\alpha \Im \left\lbrace \sigma_{V,1}^+ F_{\perp_0}F^*_{\perp_1}-\sigma_{V,1}^- F_{\para_0} F^*_{\para_1}  -4\, \sigma_{T,1} \bigg( F^T_{\perp_0}F^{T*}_{\perp_1} - F^T_{\para_0} F^{T*}_{\para_1} \bigg) \right\rbrace  \,, \\
    \label{eq:K3sm0}
    \frac{\dd\Gamma}{\dd q^2}
    K_{3s}
        & = -\frac{\alpha}{\sqrt{2}} \Im \left\lbrace \sigma_{V,2} \bigg( F_{\para_1} F^*_{\perp_0} + F_{\para_0} F^{*}_{\perp_1} \bigg)  +4\,\sigma_{ST,1}^+ F^S_{\perp} F^{T*}_{\perp_1} - 4\,\sigma_{ST,1}^- F^S_{\para} F^{T*}_{\para_1} \right\rbrace  \,, \\
    \label{eq:K4scm0}
    \frac{\dd\Gamma}{\dd q^2}
    K_{4sc} 
        & = \frac{\alpha}{\sqrt{2}} \Re \left\lbrace \sigma_{V,2} \bigg( F_{\para_1} F^*_{\perp_0} - F_{\para_0} F^*_{\perp_1} \bigg)  + 8\, \sigma_{T,1} \bigg( F^T_{\perp_0} F^{T*}_{\para_1} - F^T_{\para_0} F^{T*}_{\perp_1} \bigg) \right\rbrace \,,\\
    \label{eq:K4sm0}
    \frac{\dd\Gamma}{\dd q^2}
    K_{4s}
        & = -\sqrt{2}\, \alpha \Re \left\lbrace \sigma_{V,1}^+ F_{\perp_0} F^*_{\perp-1} - \sigma_{V,1}^- F_{\para_0} F^*_{\para_1} \right\rbrace \\
    \nonumber 
        & \quad -2\,\sqrt{2}\,\alpha \Re \left\lbrace \sigma_{ST,1}^+ F^S_{\perp} F^{T*}_{\para_1} - 2\,\sigma_{ST,1}^- F^S_{\para} F^{T*}_{\perp_1} \right\rbrace \,.
\end{align}
The result above illustrates that the cascade decay for $\ell=e,\mu$ is sensitive to 12 out of the 25 real-valued combinations
of Wilson coefficients:
\begin{equation*}
    \Sigma(m_\ell = 0) \equiv \lbrace \sigma_{V,1}^\pm, \sigma_{S,1}^\pm, \sigma_{T,1}, \Re \sigma_{V,2}, \Im \sigma_{V,2}, \Re \sigma_{S,2}, \Re \sigma_{ST,1}^\pm, \Im \sigma_{ST,1}^\pm \rbrace\,.
\end{equation*}
This sensitivity should be compared to the sensitivity exhibited by $\bar{B}\to D^{(*)}\ell^-\bar\nu$ decays.
For the pseudoscalar final state meson, the coefficients $\wilson{V,-}$ and $\wilson{S,-}$ do not enter at all by virtue of vanishing matrix elements.
For the vector final state meson, the coefficient $\wilson{S,+}$ does not enter for the same reason.
Hence, not all elements of $\Sigma(m_\ell = 0)$ are accessible in the $B$-meson decays, and therefore the $\Lambda_b$ cascade decay
provides new and complementary constraints on the Wilson coefficients.
We refrain from providing lengthy expressions for the massive case $\ell = \tau$, which can be readily obtained from our results in
\refsec{analytical-results:ang-dist}.
We find that the cascade decay with a massive lepton is sensitive to a larger set of 22 (out of 25) real-valued
combinations of Wilson coefficients. The full set of combinations reads:
\begin{equation*}
    \Sigma \equiv \Sigma(m_\ell = 0) \cup \lbrace \sigma_{VS,1}^\pm, \Re \sigma_{VS,2}, \Re \sigma_{SV,2}, \sigma_{VT,1}^\pm \rbrace\,.
\end{equation*}

Presently, the only prospect for measurement of the $\Lambda_b^0\to \Lambda_c^+\mu^-\bar\nu$ decays
is the LHCb experiment. Within LHCb, reconstruction of the $\bar\nu$ momentum from the remaining
initial state and decay kinematics is difficult, but possible \cite{Ciezarek:2016lqu}.
We do not expect a measurement of all ten angular observables before a sufficiently large data set
is available.
In the mean time, we focus on identifying observables that can be more readily measured than the full
angular distribution. Analysis of \refeq{angular-distribution} suggests only one angular observable
whose determination can be achieved without reconstruction of either the dilepton helicity angle or the azimuthal angle.
This observable is the hadronic forward-backward asymmetry,
\begin{equation}
    A_\text{FB}^{\Lambda_c} \equiv K_{2ss} + \frac{1}{2}K_{2cc}\,.
\end{equation}
Measurement of this observable can be achieved through a counting experiment with respect to
the sign of $\cos\theta_{\Lambda_c}$, normalized to the sum of all events.
The normalization through the decay rate is helpful in two ways. First, it reduces the inherent
theoretical uncertainties by partial cancellation of the (correlated) form factor uncertainties.
Second, it reduces systematic experimental uncertainties in the measurement, including the
poorly-known $\Lambda_b$ fragmentation fraction at LHCb. The ability to determine $A_\text{FB}^{\Lambda_c}$ regardless
of the reconstruction of the lepton helicity angle will also make possible to probe LFU
through the ratio:
\begin{equation}
    \label{eq:RAFBh}
    R(A_\text{FB}^{\Lambda_c}) \equiv \frac{\left[A_\text{FB}^{\Lambda_c}\right]_{\ell=\tau}}{\left[A_\text{FB}^{\Lambda_c}\right]_{\ell=\mu}}\,.
\end{equation}
In the above, the $\ell=\tau$ and $\ell=\mu$ subscripts denote that the forward-backward asymmetry is to be extracted from decays into the specific lepton species $\ell$.
By virtue of taking the LFU ratio as defined in \refeq{RAFBh} the sensitivity to the parity violating parameter $\alpha$ is completely removed. \\

For further phenomenological applications we provide the full set of angular observables for the full basis of dimension-six
operators as part of the \EOS software \cite{EOS} as of version 0.2.5.

\begin{table}[ht]
    \centering
    \begin{tabular}{l c c c}
        \toprule
        ~
            & \multicolumn{2}{c}{SM}
            & NP
            \\
        obs.
            & $\ell=\mu$
            & $\ell=\tau$
            & $\ell=\tau$
            \\
        \midrule
$K_{1cc}$ & $+0.206 \pm 0.004$ & $+0.310 \pm 0.001$ & $+0.311 \pm 0.000$\\
$K_{1c}$ & $-0.134 \pm 0.004$ & $+0.016 \pm 0.003$ & $+0.037 \pm 0.002$\\
$K_{2ss}$ & $+0.288 \pm 0.032$ & $+0.221 \pm 0.024$ & $+0.228 \pm 0.025$\\
$K_{2cc}$ & $+0.115 \pm 0.013$ & $+0.183 \pm 0.020$ & $+0.193 \pm 0.021$\\
$K_{2c}$ & $-0.164 \pm 0.018$ & $-0.031 \pm 0.004$ & $-0.017 \pm 0.003$\\
$K_{4sc}$ & $+0.063 \pm 0.008$ & $+0.023 \pm 0.003$ & $+0.022 \pm 0.002$\\
$K_{4s}$ & $+0.125 \pm 0.015$ & $+0.063 \pm 0.007$ & $+0.065 \pm 0.007$\\        
        \bottomrule
    \end{tabular}
    \caption{Predictions for the angular observables in the SM and in the NP benchmark point.
    The observable $K_{1ss}$ can be obtained as $(1 - K_{1cc})/2$ and is hence not listed.
    The observables $K_{3sc}$ and $K_{3s}$ are zero in the SM and in all NP models without new CP-violating phases in the
    $b\to c\ell\bar\nu$ Wilson coefficients.
    }
    \label{tab:Ki}
\end{table}

\section{Numerical Results}
\label{sec:numerical-results}

For the numerical illustration we define a NP benchmark point:
\begin{equation}
\begin{aligned}
    \wilson[(\tau)]{V,L} & =  1.15\,, &
    \wilson[(\tau)]{V,R} & =  0\,,    \\
    \wilson[(\tau)]{S,L} & = -0.3\,,  &
    \wilson[(\tau)]{S,R} & = +0.3\,,  &
    \wilson[(\tau)]{T}   & =  0\,.
\end{aligned}
\end{equation}
This point is inspired by and compatible with the best-fit point labelled ``minimum 1'' in a recent global
analysis of the available data on $b\to c\tau\bar\nu$ processes \cite{Murgui:2019czp}.\\

\begin{figure}[th]
    \centering
    \includegraphics[width=.69\textwidth]{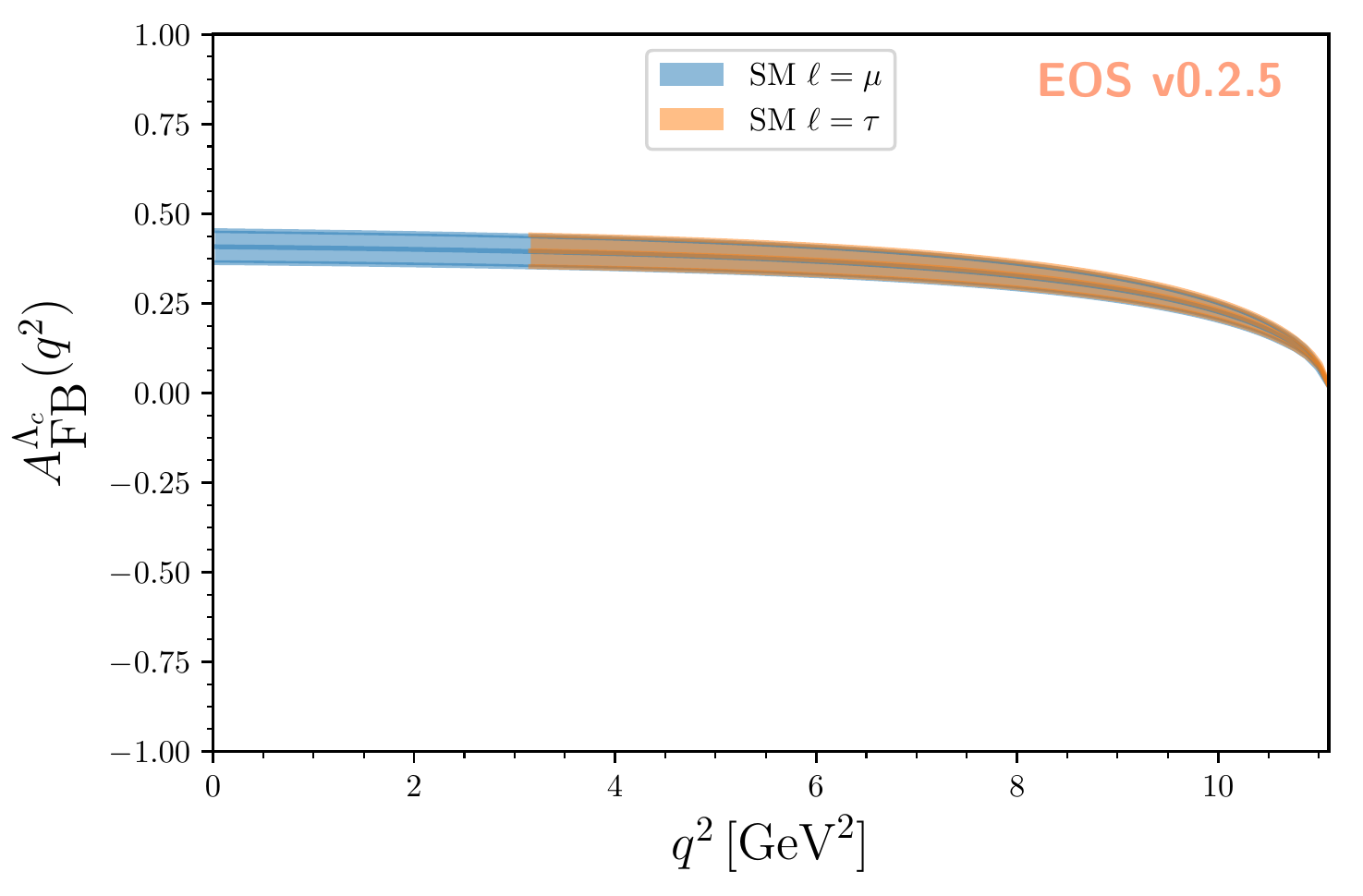}\\
    \includegraphics[width=.69\textwidth]{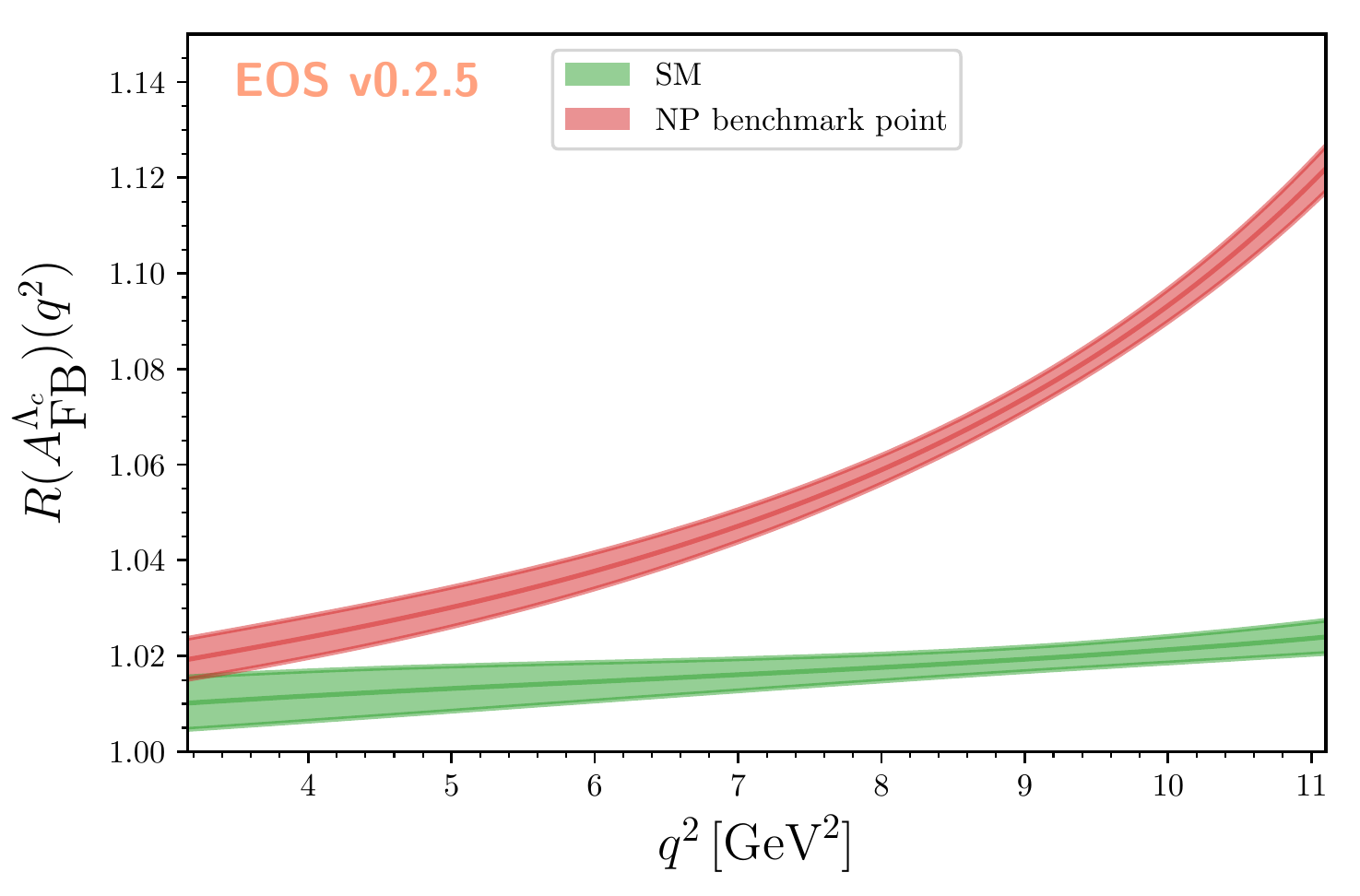}
    \caption{Predictions for (top) the hadronic forward-backward asymmetry in the SM for $\ell=\mu$ and $\ell=\tau$ final states,
    and (bottom) the LFU ratio $R(A_\text{FB}^{\Lambda_c})$ in the SM and in the NP benchmark point. The bands correspond to
    the $68\%$ probability envelopes. The theoretical uncertainties are due to the hadronic form factors \cite{Detmold:2015aaa} and the parity violating
    parameter $\alpha$ (see text). Due to the absence of any threshold effects with respect to the dilepton mass,
    the two bands for the hadronic forward-backward asymmetry are virtually indistinguishable. The LFU ratio hence exhibits this large sensitivity
    to the NP benchmark point.}
    \label{fig:AFBh}
\end{figure}

We provide numerical results for the entire set of angular observables integrated over their full $q^2$ phase space in
\reftab{Ki}. Our results include SM predictions for both $\ell=\mu$ and $\ell=\tau$, as well as predictions for the
NP benchmark point ($\ell=\tau$ only).
We illustrate the $q^2$ dependence of the hadronic forward-backward asymmetry and the NP reach of its LFU ratio
in \reffig{AFBh}. For all predictions we estimate the theoretical uncertainties based on two sets of inputs.
First, for the parity violating parameter $\alpha$---following the discussion in \refsec{analytical-results:ang-dist}---we use
an average that includes the new BESIII measurement in lieu of the world average. Second, for the hadronic form factors we use the
lattice QCD results of the full set of form factors based on the tables provided in refs.~\cite{Detmold:2015aaa,Datta:2017aue}.
We use the \EOS software \cite{EOS} for the computation of all numerical values and plots shown in this work.\\

Benefiting from the correlated results for the $\Lambda_b\to \Lambda_c$ form factors and from cancellations due to the normalization
to the decay rate, we find small uncertainties of the order of $11\%$ for $A_{\text{FB}}^{\Lambda_c}$ that are dominated by the
uncertainty in the parameter $\alpha$. This becomes much more visible in the LFU ratio where $\alpha$ cancels completely.
For that observable the relative uncertainty is reduced to $\sim 1\%$, providing an excellent opportunity for a high-precision probe
of New Physics in semileptonic $b\to c$ transitions.
We therefore encourage a sensitivity study to determine the experimental precision that the LHCb experiment can achieve for the
projected size of the upcoming data sets.

\section{Summary}
\label{sec:summary}

We present the first study of the full angular distribution of the cascade decay $\Lambda_b^0\to \Lambda_c^+(\to \Lambda^0 \pi^+) \ell^-\bar\nu$
using the complete basis of dimension-six operators, assuming only left-handed neutrinos, in the weak effective field theory for $b\to c\ell\bar\nu$ transitions. As a cross check, we reproduce
the rate of the non-cascade decay. Our findings are at variance with some scalar/vector and pseudoscalar/axialvector interference terms of the
four-differential decay rate in the literature. However, our results pass a crucial cross check for amplitudes to time-like polarized dilepton states,
which is not fulfilled by the results in the literature.\\

We express the four-differential rate through a set of ten angular observables, nine of which are independent.
The full set of angular observables is shown to be sensitive to more combinations of NP couplings than the decays
$\bar{B}\to D\ell\bar\nu$ and $\bar{B}\to D^*\ell\bar\nu$ taken together. This highlights the usefulness of the cascade decay
in constraining potential NP effects in $b\to c \ell\bar\nu$ transitions for all lepton species $\ell=e,\mu,\tau$.
For convenience, we provide computer code for the numerical evaluation of all angular observables as part of the \EOS software.\\

We suggest to measure the hadronic forward-backward asymmetry, which is the only angular observable that can be extracted
from the angular distribution without either knowledge of the lepton helicity angle or the azimuthal angle between the
decay planes. We expect good prospects for its measurement at the LHCb experiment for all three lepton species.
We find that the LFU ratio of the hadronic forward-backward asymmetry features very small hadronic uncertainties
in the Standard Model and beyond. It is therefore a prime candidate to cross check the present anomalies in the
LFU ratios $R(D)$ and $R(D^*)$.

\acknowledgments

This work is supported by the DFG within the Emmy Noether Programme under grant DY-130/1-1
and the DFG Collaborative Research Center 110 ``Symmetries and the Emergence of Structure in QCD''.

\appendix

\section{Kinematics}
\label{app:kinematics}
We label the momenta in the cascade decay as follows:
\begin{align}
    \Lambda^0_b(p) \to \Lambda_c^+(k) \bar{\nu}_\ell (q_1) \ell^-(q_2) \,, \quad \Lambda^+_c(k) \to \Lambda^0(k_1) \pi^+(k_2) \,.
\end{align}
The $z$-axis is chosen in the $\Lambda_b$ rest frame such that the $\Lambda_c$ travels in positive and the $W$ boson in negative $z$-direction.
Within the $\Lambda_c$ rest frame, we define the angle $\theta_{\Lambda_c}$ as the angle enclosed by the flight direction of the $\Lambda$ baryon and the $z$-axis.
Analogously, within the dilepton rest frame, $\theta_\ell$ is defined as the angle between the charged lepton momentum and the $z$-axis.
Finally, we define $\phi$ as the azimuthal angle of the $\pi^+$, i.e., if $\phi=0$, the coordinate system is chosen such that the charged lepton
and the $\pi^+$ both travel in positive $x$-direction.\\

The polarization vectors in the dilepton rest frame we choose to be
\begin{align}
    \epsilon^\mu(t) = (1,0,0,0)^T \,, \quad \epsilon^\mu(\pm) = \frac{1}{\sqrt{2}} (0,\pm 1,-i,0)^T \,, \quad \epsilon^\mu(0) = (0,0,0,-1)^T \,.
\end{align}
Furthermore, we define
\begin{align}
   \bar{q} = q_1-q_2 \,, \qquad  \bar{k}=k_1-k_2 \,.
\end{align}
With these definitions, some of the kinematical Lorentz invariants are
\begin{align}
    \bar{q}\cdot \epsilon(\pm) & = \pm\sqrt{\frac{q^2}{2}} \bigg( 1-\frac{m_\ell^2}{q^2} \bigg) \sin(\theta_\ell) \,,\\
    \bar{q}\cdot \epsilon(0) &= - \sqrt{q^2} \bigg(1-\frac{m_\ell^2}{q^2} \bigg) \cos(\theta_\ell)   \,, \\ 
    \epsilon_{\mu\nu\lambda\rho} \epsilon^\mu(\pm) \epsilon^\nu(\mp) q_1^\lambda q_2^\rho & = \pm \frac{i}{2} (q^2-m_\ell^2)\cos(\theta_\ell)  \,,\\
    \bar{k}\cdot\epsilon(\pm) &= \pm \sqrt{\frac{r_+ r_-}{2 m_{\Lambda_c}^2}} \sin(\theta_{\Lambda_c}) e^{\mp i \phi} \,, \\
    \bar{k}\cdot\epsilon(0) &= \frac{(m_{\Lambda}^2-m_{\pi}^2)\sqrt{s_+ s_-} + (m_{\Lambda_b}^2-m_{\Lambda_c}^2-q^2)\sqrt{r_+ r_-} \cos(\theta_{\Lambda_c})}{2 m_{\Lambda_c}^2 \sqrt{q^2}} \,, \\
    \epsilon_{\mu\nu\lambda\rho} \epsilon^\mu(\pm) \epsilon^\nu(\mp) k_1^\lambda k_2^\rho &= \mp \frac{i}{2} \sqrt{r_+ r_-} \cos(\theta_{\Lambda_c}) \,.
\end{align}
with $\epsilon^{0123} = -\epsilon_{0123} = +1$. We defined
\begin{align}
    r_\pm = (m_{\Lambda_c}^2\pm m_{\Lambda}^2 )^2 - m_\pi^2 \,, \quad \text{such that} \quad r_+r_- = \lambda(m_{\Lambda_c}^2,m_{\Lambda}^2,m_\pi^2) \,.
\end{align}

\bibliographystyle{JHEPmod}
\bibliography{references}

\providecommand{\href}[2]{#2}\begingroup\raggedright\begin{thebibliography}{10}

\bibitem{Lees:2012xj}
{\scshape BaBar} collaboration, J.~P. Lees et~al., \emph{{Evidence for an
  excess of $\bar{B} \to D^{(*)} \tau^-\bar{\nu}_\tau$ decays}},
  \href{https://doi.org/10.1103/PhysRevLett.109.101802}{\emph{Phys. Rev. Lett.}
  {\bfseries 109} (2012) 101802}
  [\href{https://arxiv.org/abs/1205.5442}{{\ttfamily 1205.5442}}].

\bibitem{Lees:2013uzd}
{\scshape BaBar} collaboration, J.~P. Lees et~al., \emph{{Measurement of an
  Excess of $\bar{B} \to D^{(*)}\tau^- \bar{\nu}_\tau$ Decays and Implications
  for Charged Higgs Bosons}},
  \href{https://doi.org/10.1103/PhysRevD.88.072012}{\emph{Phys. Rev.}
  {\bfseries D88} (2013) 072012}
  [\href{https://arxiv.org/abs/1303.0571}{{\ttfamily 1303.0571}}].

\bibitem{Huschle:2015rga}
{\scshape Belle} collaboration, M.~Huschle et~al., \emph{{Measurement of the
  branching ratio of $\bar{B} \to D^{(\ast)} \tau^- \bar{\nu}_\tau$ relative to
  $\bar{B} \to D^{(\ast)} \ell^- \bar{\nu}_\ell$ decays with hadronic tagging
  at Belle}}, \href{https://doi.org/10.1103/PhysRevD.92.072014}{\emph{Phys.
  Rev.} {\bfseries D92} (2015) 072014}
  [\href{https://arxiv.org/abs/1507.03233}{{\ttfamily 1507.03233}}].

\bibitem{Aaij:2015yra}
{\scshape LHCb} collaboration, R.~Aaij et~al., \emph{{Measurement of the ratio
  of branching fractions $\mathcal{B}(\bar{B}^0 \to
  D^{*+}\tau^{-}\bar{\nu}_{\tau})/\mathcal{B}(\bar{B}^0 \to
  D^{*+}\mu^{-}\bar{\nu}_{\mu})$}},  [Erratum: Phys. Rev.
  Lett.115,no.15,159901(2015)]\href{https://doi.org/10.1103/PhysRevLett.115.159901,
  10.1103/PhysRevLett.115.111803}{\emph{Phys. Rev. Lett.} {\bfseries 115}
  (2015) 111803} [\href{https://arxiv.org/abs/1506.08614}{{\ttfamily
  1506.08614}}].

\bibitem{Hirose:2016wfn}
{\scshape Belle} collaboration, S.~Hirose et~al., \emph{{Measurement of the
  $\tau$ lepton polarization and $R(D^*)$ in the decay $\bar{B} \to D^* \tau^-
  \bar{\nu}_\tau$}},
  \href{https://doi.org/10.1103/PhysRevLett.118.211801}{\emph{Phys. Rev. Lett.}
  {\bfseries 118} (2017) 211801}
  [\href{https://arxiv.org/abs/1612.00529}{{\ttfamily 1612.00529}}].

\bibitem{Hirose:2017dxl}
{\scshape Belle} collaboration, S.~Hirose et~al., \emph{{Measurement of the
  $\tau$ lepton polarization and $R(D^*)$ in the decay $\bar{B} \rightarrow D^*
  \tau^- \bar{\nu}_\tau$ with one-prong hadronic $\tau$ decays at Belle}},
  \href{https://doi.org/10.1103/PhysRevD.97.012004}{\emph{Phys. Rev.}
  {\bfseries D97} (2018) 012004}
  [\href{https://arxiv.org/abs/1709.00129}{{\ttfamily 1709.00129}}].

\bibitem{Aaij:2017uff}
{\scshape LHCb} collaboration, R.~Aaij et~al., \emph{{Measurement of the ratio
  of the $B^0 \to D^{*-} \tau^+ \nu_{\tau}$ and $B^0 \to D^{*-} \mu^+
  \nu_{\mu}$ branching fractions using three-prong $\tau$-lepton decays}},
  \href{https://doi.org/10.1103/PhysRevLett.120.171802}{\emph{Phys. Rev. Lett.}
  {\bfseries 120} (2018) 171802}
  [\href{https://arxiv.org/abs/1708.08856}{{\ttfamily 1708.08856}}].

\bibitem{Aaij:2017deq}
{\scshape LHCb} collaboration, R.~Aaij et~al., \emph{{Test of Lepton Flavor
  Universality by the measurement of the $B^0 \to D^{*-} \tau^+ \nu_{\tau}$
  branching fraction using three-prong $\tau$ decays}},
  \href{https://doi.org/10.1103/PhysRevD.97.072013}{\emph{Phys. Rev.}
  {\bfseries D97} (2018) 072013}
  [\href{https://arxiv.org/abs/1711.02505}{{\ttfamily 1711.02505}}].

\bibitem{Abdesselam:2019dgh}
{\scshape Belle} collaboration, A.~Abdesselam et~al., \emph{{Measurement of
  $\mathcal{R}(D)$ and $\mathcal{R}(D^{\ast})$ with a semileptonic tagging
  method}},  \href{https://arxiv.org/abs/1904.08794}{{\ttfamily 1904.08794}}.

\bibitem{Neubert:1992pn}
M.~Neubert, Z.~Ligeti and Y.~Nir, \emph{{The Subleading Isgur-Wise form-factor
  chi(3) (v, v-prime) to order alpha-s in QCD sum rules}},
  \href{https://doi.org/10.1103/PhysRevD.47.5060}{\emph{Phys. Rev.} {\bfseries
  D47} (1993) 5060} [\href{https://arxiv.org/abs/hep-ph/9212266}{{\ttfamily
  hep-ph/9212266}}].

\bibitem{Ligeti:1993hw}
Z.~Ligeti, Y.~Nir and M.~Neubert, \emph{{The Subleading Isgur-Wise form-factor
  Xi-3 (v - v-prime) and its implications for the decays $\bar{B} \to D^* \ell
  \bar\nu$}}, \href{https://doi.org/10.1103/PhysRevD.49.1302}{\emph{Phys. Rev.}
  {\bfseries D49} (1994) 1302}
  [\href{https://arxiv.org/abs/hep-ph/9305304}{{\ttfamily hep-ph/9305304}}].

\bibitem{Faller:2008tr}
S.~Faller, A.~Khodjamirian, C.~Klein and T.~Mannel, \emph{{$B \to D^{(*)}$ Form
  Factors from QCD Light-Cone Sum Rules}},
  \href{https://doi.org/10.1140/epjc/s10052-009-0968-4}{\emph{Eur. Phys. J.}
  {\bfseries C60} (2009) 603}
  [\href{https://arxiv.org/abs/0809.0222}{{\ttfamily 0809.0222}}].

\bibitem{Fajfer:2012vx}
S.~Fajfer, J.~F. Kamenik and I.~Nisandzic, \emph{{On the $B \to D^* \tau \bar
  \nu_{\tau}$ Sensitivity to New Physics}},
  \href{https://doi.org/10.1103/PhysRevD.85.094025}{\emph{Phys. Rev.}
  {\bfseries D85} (2012) 094025}
  [\href{https://arxiv.org/abs/1203.2654}{{\ttfamily 1203.2654}}].

\bibitem{Na:2015kha}
{\scshape HPQCD} collaboration, H.~Na, C.~M. Bouchard, G.~P. Lepage, C.~Monahan
  and J.~Shigemitsu, \emph{{$B \rightarrow D \ell \bar\nu$ form factors at
  nonzero recoil and extraction of $|V_{cb}|$}},  [Erratum: Phys.
  Rev.D93,no.11,119906(2016)]\href{https://doi.org/10.1103/PhysRevD.93.119906,
  10.1103/PhysRevD.92.054510}{\emph{Phys. Rev.} {\bfseries D92} (2015) 054510}
  [\href{https://arxiv.org/abs/1505.03925}{{\ttfamily 1505.03925}}].

\bibitem{Lattice:2015rga}
{\scshape MILC} collaboration, J.~A. Bailey et~al., \emph{{$B\to D\ell\bar\nu$
  form factors at nonzero recoil and $|V_{cb}|$ from 2+1-flavor lattice QCD}},
  \href{https://doi.org/10.1103/PhysRevD.92.034506}{\emph{Phys. Rev.}
  {\bfseries D92} (2015) 034506}
  [\href{https://arxiv.org/abs/1503.07237}{{\ttfamily 1503.07237}}].

\bibitem{Neubert:1992wq}
M.~Neubert, Z.~Ligeti and Y.~Nir, \emph{{QCD sum rule analysis of the
  subleading Isgur-Wise form-factor Chi-2 (v v-prime)}},
  \href{https://doi.org/10.1016/0370-2693(93)90728-Z}{\emph{Phys. Lett.}
  {\bfseries B301} (1993) 101}
  [\href{https://arxiv.org/abs/hep-ph/9209271}{{\ttfamily hep-ph/9209271}}].

\bibitem{Wang:2017jow}
Y.-M. Wang, Y.-B. Wei, Y.-L. Shen and C.-D. Lü, \emph{{Perturbative
  corrections to $B \to D$ form factors in QCD}},
  \href{https://doi.org/10.1007/JHEP06(2017)062}{\emph{JHEP} {\bfseries 06}
  (2017) 062} [\href{https://arxiv.org/abs/1701.06810}{{\ttfamily
  1701.06810}}].

\bibitem{Gubernari:2018wyi}
N.~Gubernari, A.~Kokulu and D.~van Dyk, \emph{{$B\to P$ and $B\to V$ Form
  Factors from $B$-Meson Light-Cone Sum Rules beyond Leading Twist}},
  \href{https://doi.org/10.1007/JHEP01(2019)150}{\emph{JHEP} {\bfseries 01}
  (2019) 150} [\href{https://arxiv.org/abs/1811.00983}{{\ttfamily
  1811.00983}}].

\bibitem{Datta:2017aue}
A.~Datta, S.~Kamali, S.~Meinel and A.~Rashed, \emph{{Phenomenology of $
  {\Lambda}_b\to {\Lambda}_c\tau {\overline{\nu}}_{\tau } $ using lattice QCD
  calculations}}, \href{https://doi.org/10.1007/JHEP08(2017)131}{\emph{JHEP}
  {\bfseries 08} (2017) 131}
  [\href{https://arxiv.org/abs/1702.02243}{{\ttfamily 1702.02243}}].

\bibitem{Shivashankara:2015cta}
S.~Shivashankara, W.~Wu and A.~Datta, \emph{{$\Lambda_b \to \Lambda_c \tau
  \bar{\nu}_{\tau}$ Decay in the Standard Model and with New Physics}},
  \href{https://doi.org/10.1103/PhysRevD.91.115003}{\emph{Phys. Rev.}
  {\bfseries D91} (2015) 115003}
  [\href{https://arxiv.org/abs/1502.07230}{{\ttfamily 1502.07230}}].

\bibitem{Gutsche:2015mxa}
T.~Gutsche, M.~A. Ivanov, J.~G. Körner, V.~E. Lyubovitskij, P.~Santorelli and
  N.~Habyl, \emph{{Semileptonic decay $\Lambda_b \to \Lambda_c + \tau^- +
  \bar{\nu_\tau}$ in the covariant confined quark model}},  [Erratum: Phys.
  Rev.D91,no.11,119907(2015)]\href{https://doi.org/10.1103/PhysRevD.91.074001,
  10.1103/PhysRevD.91.119907}{\emph{Phys. Rev.} {\bfseries D91} (2015) 074001}
  [\href{https://arxiv.org/abs/1502.04864}{{\ttfamily 1502.04864}}].

\bibitem{Boer:2014kda}
P.~Böer, T.~Feldmann and D.~van Dyk, \emph{{Angular Analysis of the Decay
  $\Lambda_b \to \Lambda (\to N \pi) \ell^+\ell^-$}},
  \href{https://doi.org/10.1007/JHEP01(2015)155}{\emph{JHEP} {\bfseries 01}
  (2015) 155} [\href{https://arxiv.org/abs/1410.2115}{{\ttfamily 1410.2115}}].

\bibitem{Das:2018sms}
D.~Das, \emph{{Model independent New Physics analysis in
  $\Lambda_b\to\Lambda\mu^+\mu^-$ decay}},
  \href{https://doi.org/10.1140/epjc/s10052-018-5731-2}{\emph{Eur. Phys. J.}
  {\bfseries C78} (2018) 230}
  [\href{https://arxiv.org/abs/1802.09404}{{\ttfamily 1802.09404}}].

\bibitem{Das:2018iap}
D.~Das, \emph{{On the angular distribution of $\Lambda_b\to\Lambda(\to
  N\pi)\tau^+\tau^-$ decay}},
  \href{https://doi.org/10.1007/JHEP07(2018)063}{\emph{JHEP} {\bfseries 07}
  (2018) 063} [\href{https://arxiv.org/abs/1804.08527}{{\ttfamily
  1804.08527}}].

\bibitem{Meinel:2016dqj}
S.~Meinel, \emph{{$\Lambda_c \to \Lambda l^+ \nu_l$ form factors and decay
  rates from lattice QCD with physical quark masses}},
  \href{https://doi.org/10.1103/PhysRevLett.118.082001}{\emph{Phys. Rev. Lett.}
  {\bfseries 118} (2017) 082001}
  [\href{https://arxiv.org/abs/1611.09696}{{\ttfamily 1611.09696}}].

\bibitem{Ablikim:2019zwe}
{\scshape BESIII} collaboration, M.~Ablikim et~al., \emph{{Measurements of Weak
  Decay Asymmetries of $\Lambda_c^+\to pK_S^0$, $\Lambda\pi^+$,
  $\Sigma^+\pi^0$, and $\Sigma^0\pi^+$}},
  \href{https://arxiv.org/abs/1905.04707}{{\ttfamily 1905.04707}}.

\bibitem{Descotes-Genon:2018foz}
S.~Descotes-Genon, A.~Falkowski, M.~Fedele, M.~González-Alonso and J.~Virto,
  \emph{{The CKM parameters in the SMEFT}},
  \href{https://doi.org/10.1007/JHEP05(2019)172}{\emph{JHEP} {\bfseries 05}
  (2019) 172} [\href{https://arxiv.org/abs/1812.08163}{{\ttfamily
  1812.08163}}].

\bibitem{Tanabashi:2018oca}
{\scshape Particle Data Group} collaboration, M.~Tanabashi et~al.,
  \emph{{Review of Particle Physics}},
  \href{https://doi.org/10.1103/PhysRevD.98.030001}{\emph{Phys. Rev.}
  {\bfseries D98} (2018) 030001}.

\bibitem{Avery:1990ya}
{\scshape CLEO} collaboration, P.~Avery et~al., \emph{{Measurement of the
  Lambda(c) decay asymmetry parameter}},
  \href{https://doi.org/10.1103/PhysRevLett.65.2842}{\emph{Phys. Rev. Lett.}
  {\bfseries 65} (1990) 2842}.

\bibitem{Albrecht:1991vs}
{\scshape ARGUS} collaboration, H.~Albrecht et~al., \emph{{A Measurement of
  asymmetry in the decay $\Lambda_c^+ \to \Lambda \pi^+$}},
  \href{https://doi.org/10.1016/0370-2693(92)90529-D}{\emph{Phys. Lett.}
  {\bfseries B274} (1992) 239}.

\bibitem{Bishai:1995gp}
{\scshape CLEO} collaboration, M.~Bishai et~al., \emph{{Measurement of the
  decay asymmetry parameters in $\Lambda_c^+ \to \Lambda \pi^+$ and
  $\Lambda_c^+ \to \Sigma^+ pi^0$}},
  \href{https://doi.org/10.1016/0370-2693(95)00280-X}{\emph{Phys. Lett.}
  {\bfseries B350} (1995) 256}
  [\href{https://arxiv.org/abs/hep-ex/9502004}{{\ttfamily hep-ex/9502004}}].

\bibitem{Link:2005ft}
{\scshape FOCUS} collaboration, J.~M. Link et~al., \emph{{Study of the decay
  asymmetry parameter and CP violation parameter in the $\Lambda_c^+ \to
  \Lambda \pi^+$ decay}},
  \href{https://doi.org/10.1016/j.physletb.2006.01.017}{\emph{Phys. Lett.}
  {\bfseries B634} (2006) 165}
  [\href{https://arxiv.org/abs/hep-ex/0509042}{{\ttfamily hep-ex/0509042}}].

\bibitem{Ciezarek:2016lqu}
G.~Ciezarek, A.~Lupato, M.~Rotondo and M.~Vesterinen, \emph{{Reconstruction of
  semileptonically decaying beauty hadrons produced in high energy pp
  collisions}}, \href{https://doi.org/10.1007/JHEP02(2017)021}{\emph{JHEP}
  {\bfseries 02} (2017) 021}
  [\href{https://arxiv.org/abs/1611.08522}{{\ttfamily 1611.08522}}].

\bibitem{EOS}
D.~van Dyk et~al., ``{EOS --- A HEP program for Flavor Observables}.''

\bibitem{Murgui:2019czp}
C.~Murgui, A.~Peñuelas, M.~Jung and A.~Pich, \emph{{Global fit to $b \to c
  \tau \nu$ transitions}},  \href{https://arxiv.org/abs/1904.09311}{{\ttfamily
  1904.09311}}.

\bibitem{Detmold:2015aaa}
W.~Detmold, C.~Lehner and S.~Meinel, \emph{{$\Lambda_b \to p \ell^-
  \bar{\nu}_\ell$ and $\Lambda_b \to \Lambda_c \ell^- \bar{\nu}_\ell$ form
  factors from lattice QCD with relativistic heavy quarks}},
  \href{https://doi.org/10.1103/PhysRevD.92.034503}{\emph{Phys. Rev.}
  {\bfseries D92} (2015) 034503}
  [\href{https://arxiv.org/abs/1503.01421}{{\ttfamily 1503.01421}}].

\end{thebibliography}\endgroup

\end{document}